\documentclass{aa}
\usepackage{graphicx}
\begin{document}
   \title{The abundance of $^{36}$S in IRC+10216 and its production in the Galaxy}
   \titlerunning{$^{36}$S in IRC+10216}
   \author{R. Mauersberger\inst{1} \and U. Ott\inst{2}
\and C. Henkel\inst{3}  \and J. Cernicharo\inst{4} \and R.
Gallino\inst{5,}\inst{6}
          }\offprints{R. Mauersberger, mauers@iram.es}
 \institute{Instituto de Radioastronom\'{\i}a Milim\'etrica, Avda.
   Divina Pastora 7, Local\,20, E-18012 Granada, Spain
  \and
Max-Planck-Institut f\"ur Chemie,  Becherweg 27, D-55128 Mainz,
Germany \and
        Max-Planck-Institut f\"ur Radioastronomie, Auf dem H\"ugel 69,
        D-53121 Bonn, Germany \and Instituto de la Estructura de la Materia, Dept. de
Astronom\'{\i}a Molecular e Infrarroja, Serrano 113, E-28006
Madrid, Spain \and Dipartimento di Fisica Generale
dell'Universit\`a di Torino, Via Pietro Giuria 1, I-10125 Torino,
Italy \and Centre for Stellar and Planetary Sciences, Monash
University, Melbourne 3800, Australia
        }
 \date{Received  ; accepted 21.6.2004}

  \abstract{The $J=2-1$ and $3-2$ rotational lines of the rare
   isotopomer C$^{36}$S and the $J=5-4$ and $6-5$ transitions of Si$^{36}$S were detected in the carbon
   star IRC+10216 (CW\,Leo). These are the first detections of $^{36}$S bearing molecules in a star and the first spectroscopic
   detection of Si$^{36}$S. From a comparison  of  $^{34}$S and $^{36}$S bearing
isotopomers,  the
   $^{34}$S/$^{36}$S isotopic ratio is 107($\pm 15$). This value is  comparable to values
   in the interstellar medium of the inner Galactic disk (115)
   but is smaller than the solar value of 288 (Ding et al. \cite{ding2001}). The increase of the
   $^{36}$S abundance relative to $^{34}$S only qualitatively follows
   model predictions of a low mass AGB star. Quantitative agreement of the observed $^{34}$S/$^{36}$S ratio
   with the stellar models can be reached if the age of IRC+10216 and Galactic chemical evolution are taken into
   account. Other less likely possibilities are the presence of
   considerable inhomogeneities
   in the interstellar medium and either
   IRC+10216 or the Sun started with a peculiar $^{36}$S abundance. Other production mechanisms potentially
   capable of enhancing the Galactic interstellar medium are
   discussed. From the observed line density toward IRC+10216 and toward Galactic star forming regions,
   we estimate the confusion limit toward those sources.
   \\\keywords{nuclear reactions,
nucleosynthesis,
      abundances -- stars: abundances -- stars: AGB and post-AGB-stars:
      individual: IRC+10216 -- ISM: abundances -- radiolines: stars}}

   \maketitle
%

\section{Introduction}
Heavy elements are formed in stars of medium or high mass. At the
end of their lifetimes, these stars recycle, via winds or via
explosions, a fraction of these metals back into the interstellar
medium (ISM). Over the aeons, interstellar abundances of such
heavy elements have been increasing and isotopic compositions have
been changing, as can be seen by comparing the composition of the
present day ISM with that of Damped Lyman-$\alpha$ systems (Rauch
\cite{rauch98}). Not only elemental but also isotopic abundances,
which are mainly determined in the radio range, are powerful tools
to investigate stellar production sites  and the star forming
history of galaxies (Wilson \& Matteucci \cite{wilson92}; Henkel
\& Mauersberger \cite{henkel93}; Henkel et al. \cite{henkel94};
Wilson \& Rood \cite{wilson94}; Kahane \cite{kahane95}; Prantzos
et al. \cite{prantzos96}; Bieging \cite{bieging97}; Henkel et al.
\cite{henkel98}).

Being one of the ten most abundant elements and possessing four
stable isotopes, sulfur is of particular interest for such
studies. Relative abundances of  stable sulfur isotopes have been
determined for meteorites (e.g. Gao \& Thiemens \cite{gao1991}),
the Moon (Thode \& Rees \cite{thode71}), the Galactic ISM (e.g.
Chin et al. \cite{chin96}), Cosmic Rays (Thayer \cite{thayer97})
and late type stars (e.g. Kahane et al. \cite{kahane88}). Even in
external galaxies $^{32}$S/$^{34}$S ratios could be measured
(Mauersberger \& Henkel \cite{mauersberger89}; Johansson et al.
\cite{johansson94}). Small isotopic variations of sulfur in
terrestrial, meteoric and planetary samples have been proposed to
be powerful indicators of chemical, geophysical and biological
processes (Canfield \cite{Canfield01}; Farquhar \& Wing
\cite{farquhar2003})

Mauersberger et al. (\cite{mauersberger96}) presented the first
interstellar detections of a $^{36}$S bearing molecule toward a
number of Galactic hot cores. While the solar $^{34}$S/$^{36}$S
ratio is 288, an interstellar ratio of 115($\pm 17$) was found.
This supported the notion that unlike other S isotopes, $^{36}$S
is a secondary-like nucleus, predominantly synthesized via the
s-process in massive stars. Most interstellar $^{34}$S/$^{36}$S
ratios were determined, however, for sources with galactocentric
distances $R_{\rm GC}\le 7$\,kpc. A positive $^{34}$S/$^{36}$S
gradient with increasing $R_{\rm GC}$, i.e. a ratio between 115
and $\sim 180$ in the solar neighborhood, is not completely ruled
out by the data of Mauersberger et al. (\cite{mauersberger96}).

In order to further constrain the production site of $^{36}$S  in
the Galaxy, we conducted a search for rotational lines of
C$^{36}$S and Si$^{36}$S toward the prototypical carbon star
IRC+10216, where  all other stable S isotopes  have been detected
previously (Kahane et al. \cite{kahane88}).
\begin{figure}
\includegraphics[width=8.5cm]{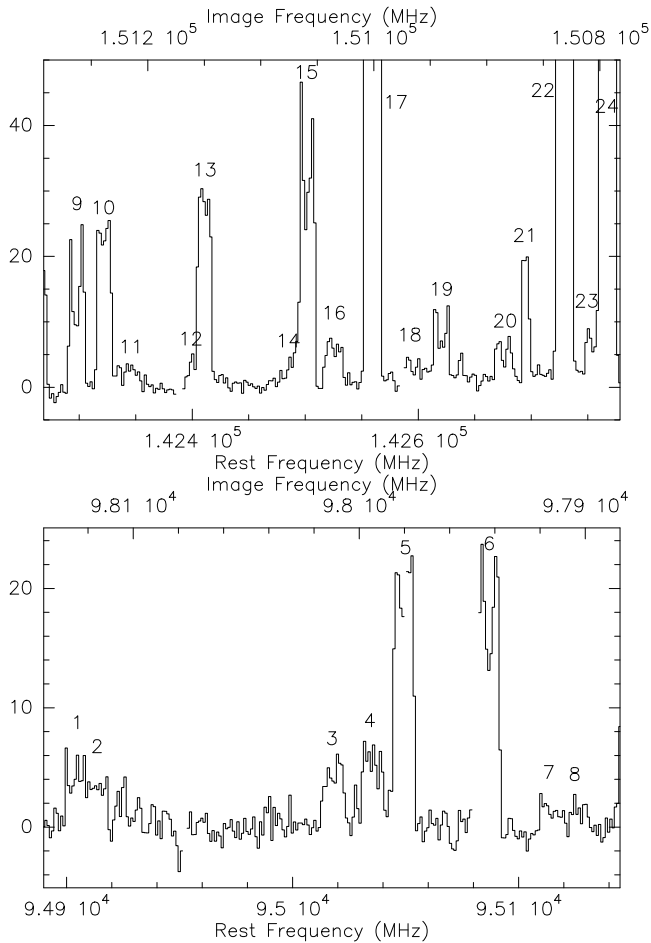}
\caption[]{The whole spectral range measured near the C$^{36}$S
$J=2-1$ (lower panel) and $3-2$ (upper panel) transitions toward
IRC+10216. The channel spacing corresponds to $3.1\,\rm
km\,s^{-1}$ for the 3\,mm spectrum and $4.2\,\rm km\,s^{-1}$ for
the 2\,mm spectrum. The intensity is in mK ($T_{\rm A}^*$).
\label{spectraall} For identifications, see the
Appendix\,\ref{appa}. }
\end{figure}

\section{Line frequencies}
The rest frequencies of the C$^{36}$S $J=2-1$ and $3-2$ lines are
95.0167 and 142.5228 GHz (V. Ahrens, pers. comm.), and 96.4130 and
144.6171 GHz for the corresponding C$^{34}$S lines (Lovas
\cite{Lovas92}).

The frequencies of Si$^{36}$S have not been measured in the
laboratory. However, it is straightforward to derive the
rotational constants of Si$^{36}$S from those of SiS (see, e.g.,
Frum et al. \cite{Frum}) using the isotopic relation for diatomic
molecules (e.g. Townes \& Schawlow \cite{townesandschawlow}). For
Si$^{36}$S, we derived $B_{0}$=8607.47\,MHz and
$D_{0}$=5.4\,10$^{-3}$\,MHz. The computed frequencies are
$\nu(J=5-4)$=86071.984\,MHz and $\nu(J=6-5)$=103284.96\,MHz. The
expected relative error for the rotational constants is $<10^{-5}$
and the resulting expected uncertainty for the $J=5-4$ and $6-5$
lines of Si$^{36}$S is $<0.5$\,MHz. We have checked the precision
of those relations for the different isotopes of SiS that have
been measured in the laboratory and we find that the quoted
uncertainties are rather conservative. Recently, Sanz et al.
(\cite{Sanz}) measured the rotational constants of several
isotopes of SiS and have provided a fit, taking into account the
breakdown of the Born-Oppenheimer approximation, to the data of
all isotopes of SiS. From their constants  we infer those of
Si$^{36}$S to be $B_{0}$=8607.495\,MHz,
$D_{0}$=5.364\,10$^{-3}$\,MHz, $\nu(J=5-4)$= 86072.27\,MHz and
$\nu(J=6-5)$=103285.31\,MHz, i.e, 0.3 MHz of difference with
respect to our early calculations. Hence, we are confident that
the calculated frequencies have an error well below 1 MHz.
\section{Observations}
The $J=2-1$ and $J=3-2$ lines of C$^{34}$S and C$^{36}$S were
observed between October 1999 and September 2003  with the IRAM
30\,m radiotelescope on Pico Veleta (Southern Spain) toward
IRC+10216. The Si$^{36}$S observations were carried out between
1996 and 2003 at the same telescope. Several spectra at different
frequency settings were observed for each line of Si$^{36}$S to
ensure that the observed lines are coming from the signal side
band.  The observations were conducted under good weather
conditions. SIS receivers were used with image sideband rejections
of 20---25\,dB. An antenna temperature ($T_{\rm A}^{*}$) scale was
established by a chopper wheel method. The forward efficiency of
the telescope was 0.94 and the beam efficiencies were 0.80 at
3\,mm and 0.65 at 2\,mm.  The beamwidth (full width at half
maximum) of the 30\,m antenna was 26$''$ and 17$''$ for the $2-1$
and $3-2$ transitions of the isotopomers of CS; for the $5-4$ and
$6-5$ transitions of SiS, beamwidths were 28$''$ and 24$''$. From
scans of continuum sources, we estimate the pointing to be correct
within 5$''$.

As backends, we employed filterbanks with 256 channels and a
channel spacing of 1\,MHz corresponding to $\sim 3.1\,\rm
km\,s^{-1}$ for the 3\,mm transitions and 2.1 km\,s$^{-1}$ for the
2\,mm transitions. Two receivers were employed simultaneously,
either observing orthogonal polarizations at the same wavelength
or measuring one line at 2\,mm and the other at 3\,mm using just
one polarization for each wavelength.

All spectral lines were measured using the wobbling secondary
mirror with a beam throw of 200$''$ in azimuth. The phase time was
2 seconds and the on-source integration of each subscan was 30
seconds. Subscans obtained by wobbling to the left and to the
right of our source were added to eliminate baseline ripples
caused by the asymmetry in the beam path (symmetric switching).
\section{Results}
\label{results}
\begin{figure}
\includegraphics[width=7cm]{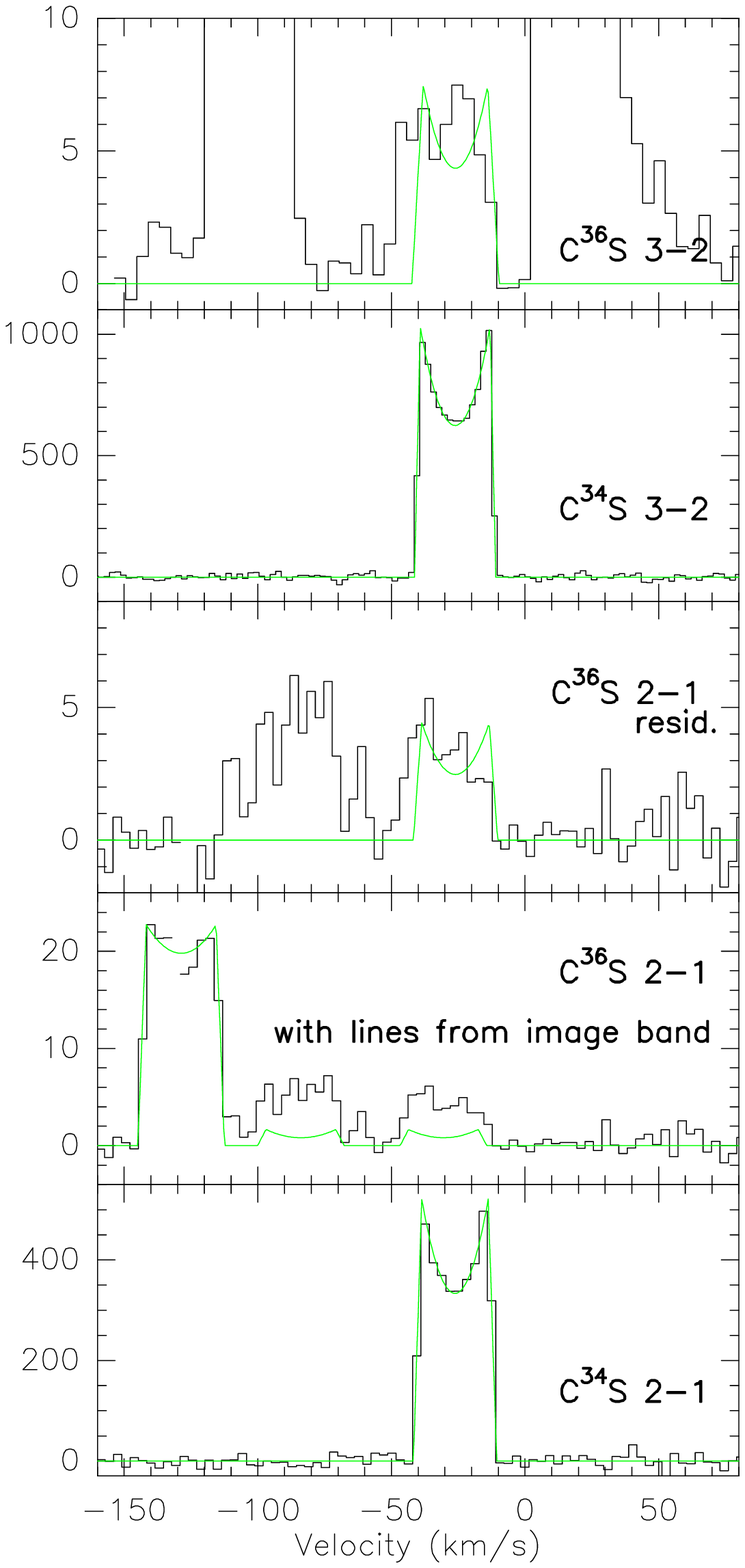}\vspace*{0.5cm}
\caption[]{\label{c36s} C$^{34}$S and C$^{36}$S  $J=2-1$ and $3-2$
spectra toward IRC+10216.  The channel spacing corresponds to
$3\,\rm km\,s^{-1}$ for the $2-1$ lines of C$^{34}$S and
C$^{36}$S, $2\,\rm km\,s^{-1}$ for the $3-2$ line of C$^{34}$S and
$4\,\rm km\,s^{-1}$ for the $3-2$ transition of C$^{36}$S.
Intensities are in mK ($T_{\rm A}^*$). Also shown are fits
(convolved with the channel widths) of a shell type emission
profile to the lines. The lineshapes of the C$^{36}$S transitions
were determined from the fits to the corresponding C$^{34}$S
lines. In the frame for C$^{36}$S $(2-1)$, the emission feature at
the extreme left corresponds to the $2-1$ line of the main isotope
of CS observed in the image band. A fit to the C$^{32}$S line and
to two l-C$_3$H lines from the image band fixing the intensity
ratios to the values obtained from a well calibrated spectrum
(Mauersberger et al. \cite{mauersberger96}) is overlayed on the
spectrum. In the third frame the residuals obtained from
subtracting the fit to the lines from the image sideband are
shown.
 \label{spectra}}
\end{figure}
\begin{figure}
\includegraphics[width=7cm]{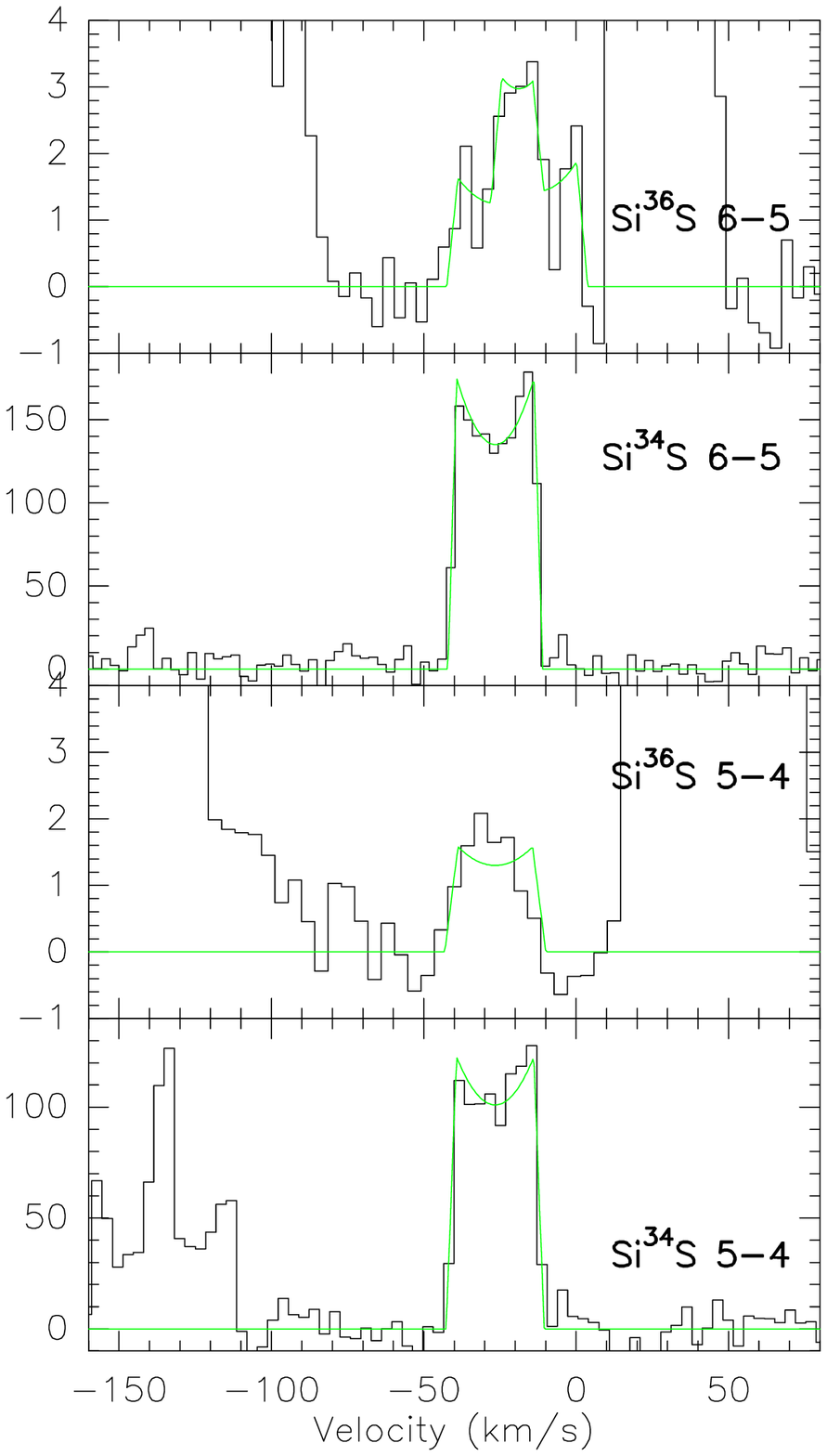}\vspace*{0.5cm}
\caption[]{Si$^{34}$S and Si$^{36}$S  $J=5-4$ and $6-5$ spectra
toward IRC+10216.  The channel spacing corresponds to $\sim$3.5
--- 4 km\,s$^{-1}$. Intensities are in mK ($T_{\rm A}^*$). Also
shown are fits (convolved with the channel widths) of a shell type
emission profile to the lines. In case of the Si$^{36}$S  $J=6-5$
transition, a simultaneous fit (assuming identical line width)
with a blended unidentified line is shown. \label{SiSspectra}}
\end{figure}

Fig. \ref{spectraall}  displays the entire spectral range measured
at the frequencies of the C$^{36}$S lines. Figs.\,\ref{c36s} and
\ref{SiSspectra} show the C$^{34}$S, C$^{36}$S, Si$^{34}$S and
Si$^{36}$S spectra that are discussed in Sect.
\ref{identification} in more detail. An inspection of
Fig.\,\ref{spectraall} shows that a main source of uncertainty in
the identification of weak lines and also in the definition of
spectral baselines comes from the presence of many weak line
features, in particular at 2\,mm. In Appendix\,\ref{appa}, we list
the line parameters of all features together with possible
identifications. The Si$^{36}$S data, showing a statistical
behavior that is similar to the C$^{36}$S data, will be analyzed
elsewhere (Cernicharo et al., in prep.).
\begin{figure}
\includegraphics[width=5.5cm,angle=270]{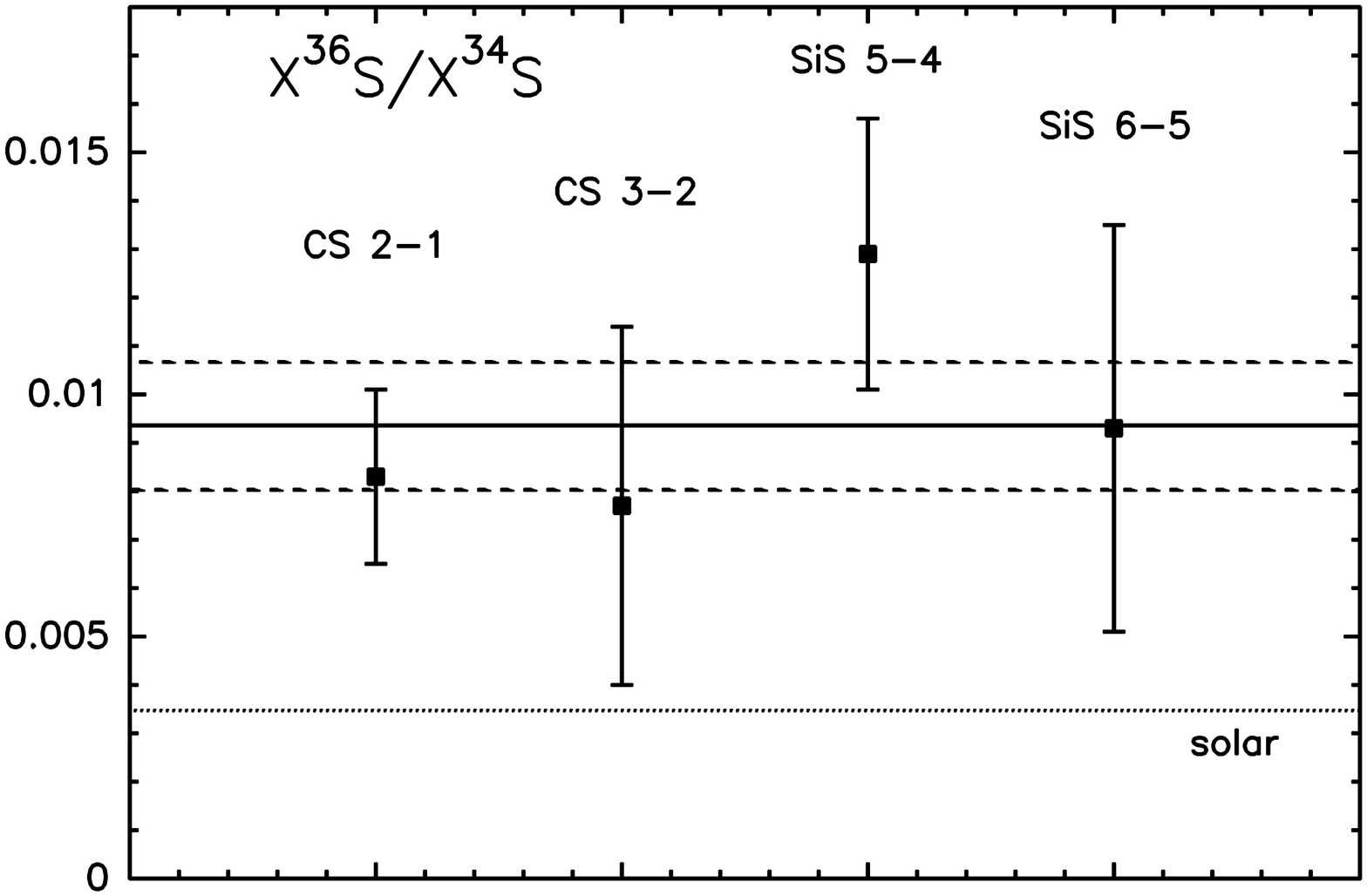}
\caption[]{\label{c34ss36sfit} Isotopic ratios of
$^{36}$S/$^{34}$S obtained from four transitions. The straight
line indicates the weighted mean value; a dashed line indicates
the error of the mean value. Also shown is the solar system ratio
(3.47\,10$^{-3}$).}
\end{figure}

From the spectra, we subtracted baselines of first order. For the
$3-2$ line of  C$^{36}$S, almost the entire frequency range is
covered with line features near or above the detection threshold
making the identification of a region free of emission a difficult
task. We chose spectral regions with particularly low emission for
the definition of a baseline.
\begin{table}
\caption[]{\label{table} Line parameters }
\begin{tabular}{r  r r r r }
\hline Transition& $\int T_{\rm A}^{*}{\rm d}v$ &$v_{\rm LSR}$ &
$v_{\rm
exp}$&$\frac{\rm C}{\rm H}$ $^{a}$\\
& mK\,km\,s$^{-1}$ &km\,s$^{-1}$&km\,s$^{-1}$&\\
\hline CS $J=2-1$\\
C$^{34}$S & 11600(100)&$-$26.2&14.0&0.55\\
C$^{36}$S & 96(21)& $-$26.2$^{b}$&14.0$^{b}$&0.55$^{b}$\\
C$^{36}$S/C$^{34}$S&\multicolumn{2}{l}{8.3(1.8)\,10$^{-3}$ $^d$}\\
\hline CS $J=3-2$\\
C$^{34}$S&22190(93)&$-$26.0&14.5&0.64\\
C$^{36}$S &170(81)&$-$26.0$^{b}$&14.5$^{b}$&0.64$^{b}$\\
C$^{36}$S/C$^{34}$S &\multicolumn{2}{l}{7.7(3.7)\,10$^{-3}$ $^d$}\\
\hline SiS $J=5-4$ \\
Si$^{34}$S & 3180(80)&$-$26.4 &14.4&0.82 \\
Si$^{36}$S & 41(9)&$-$26.4$^b$ &14.4$^b$ & 0.82$^b$\\
Si$^{36}$S/Si$^{34}$S &\multicolumn{2}{l}{12.9(2.8)\,10$^{-3}$ $^d$} \\
\hline SiS $J=6-5$\\
Si$^{34}$S & 4280(70)& $-26.5$&14.1 &0.76 \\
Si$^{36}$S$^c$ & 40(18)& $-$26.5$^b$& 14.1$^b$& 0.76$^b$\\
Si$^{36}$S/Si$^{34}$S &\multicolumn{2}{l}{9.3(4.2)\,10$^{-3}$ $^d$}\\
\hline
\end{tabular}
\\
a) the ratio between the intensities at the central velocity of
the line and at the velocity of the horns (i.e. maximum blue or
red shifted emission); b) fixed to values obtained for the
C$^{34}$S or Si$^{34}$S lines; c) simultaneous fit with a blended,
unidentified line at $\nu=103.2804$\,GHz and an intensity of
46(18)\,K\,km\,s$^{-1}$; d) from the integrated intensities
yielded by the fits to the lines of $^{34}$S and $^{36}$S bearing
species. The error given is the 1\,$\sigma$ error from the fit and
does not contain an estimate of systematic errors, e.g. due to
baseline uncertainties.
\end{table}
\section{Discussion}
\subsection{Where is the confusion limit toward IRC+10216?}
\label{confusionlim} As it can be seen in Fig.\,\ref{spectraall}
showing the total spectral range of 2$\times$256\,MHz toward
IRC+10216, we have detected a large number of spectral line
features, many of them unidentified. There are more lines at 2\,mm
than at 3\,mm although the rms of the data is similar. The
spectral line density in our 2\,mm spectrum is 15 lines per
256\,MHz. Since each line has a width of about 14.5\,MHz, about
85\% of the spectral range observed would be covered with
detectable line emission if there were no overlap. The weakest
feature we identified in Table\,\ref{lineidentification} at
$\lambda$=2\,mm has $T_{\rm A}^*$=3.5\,mK, although the rms noise
of our 2\,mm spectrum is about 0.5\,mK for 4 km\,s$^{-1}$ wide
velocity channels. Apparently lines weaker than about 3.5\,mK have
escaped detection not because of lacking signal-to-noise but
because their number is so high that they are all blended and
therefore cannot be easily distinguished.

At $\lambda=3$\,mm, 6 lines have been detected from the signal
band in a 256\,MHz wide range. Since each line has a width of
9\,MHz, 21\% of the spectral range observed is covered by detected
lines. The weakest of those have a $T_{\rm A}^*$ of about 1.5\,mK,
which corresponds to the 3$\sigma$ noise level at a velocity
resolution of 10\,km\,s$^{-1}$. In contrast to the situation at
2\,mm, we have not yet reached the 3\,mm confusion limit.
Nevertheless, at 3\,mm we also see signs of weak blended lines at
a level of $\sim$1\,mK. Weaker lines, on the order of 0.5\,mK,
have been identified recently (Cernicharo et al. \cite{cerni2004})
making use of the characteristic line shapes of IRC+10216.

To summarize, at the IRAM 30-m telescope the confusion limit
toward IRC+10216 is reached at $\lambda$=2\,mm after integrating
down to a system noise of about 1\,mK, while at 3\,mm one has to
integrate down to 0.3\,mK. For other well known molecular sources,
confusion limits are given in Appendix\,\ref{appb}.
\subsection{The identification of C$^{36}$S and Si$^{36}$S}
\label{identification} As already mentioned in
Sect.\,\ref{results} there are emission features at the rest
frequencies of both the $3-2$ and the $2-1$ lines of C$^{36}$S.
These clearly peak above the limits of line confusion. It turns
out, however, that at 3\,mm there are two emission lines of linear
C$_3$H from the image sideband, one of which coincides with the
location of the expected C$^{36}$S $2-1$ feature. Also in the band
is the image of the strong CS $2-1$ line. From a well calibrated
spectrum of these three lines (Mauersberger et al.
\cite{mauetal89}) we obtained their relative intensities.
According to Cernicharo et al. (\cite{cerni2000}) there is no hint
for variability in time, except for some very highly excited lines
or maser lines. In our spectrum of C$^{36}$S $2-1$
(Fig.\,\ref{spectra}) we could therefore apply a fit to the CS
line and the two C$_3$H lines from the image band fixing the line
shape and the relative intensities. When removing this composite
fit from the spectrum, the residual still shows emission at the
expected frequency of C$^{36}$S (Fig.\,\ref{spectra}). This
residual spectrum was used to determine the isotopic ratios given
below.

We cannot exclude that the residual line is contaminated since a
fit with fixed line shape and rest frequency to the line feature
(to the values obtained for C$^{34}$S) shows some excess emission
toward lower velocities. A possible contaminant of the $2-1$ line
of C$^{36}$S is a series of $K$ lines of the $J=23-22$ transition
of CH$_3$CCCN (methyl cyanoacetylene; Lovas \cite{Lovas84}), which
were detected toward a number of Galactic hot cores (Mauersberger
et al. \cite{mauersberger96}). This is, however, unlikely, since
this molecule is not yet identified toward IRC+10216 (e.g.
Cernicharo et al.  \cite{cerni2000}).

We have used the procedure from the CLASS data reduction package
to fit shell type circumstellar lines. In order to obtain more
reliable values for the integrated intensities, we fixed some of
the line parameters of $^{36}$S bearing species to the values
obtained from the more intense $^{34}$S bearing isotopomers. Fit
results are given in Table\,\ref{table}. In the fit to the
C$^{36}$S $3-2$ line feature there is a hint for excess emission
toward lower velocities. At that frequency, the only previously
observed interstellar line in the Lovas (\cite{Lovas84}) catalog
is a transition of CH$_2$CHCN (vinyl cyanide, $\nu = 142.523$
GHz). This complex molecule has never been detected before in
IRC+10216 and the line strength is very low. It seems plausible
that if there is contamination it comes from a transition of a
molecule abundant in late type carbon stars but not in the ISM,
such as C$_7$H and/or H$_2$C$_6$ (Gu\'elin et al. \cite{guelin
97}).

The Si$^{36}$S $5-4$ line can be well fitted, with the
determination of the baseline being the major source of
uncertainty.  The Si$^{36}$S $6-5$ line is also clearly detected,
although it seems to be blended with a weak, unidentified feature
(Fig.\,\ref{SiSspectra}). The parameters of the C$^{34}$S,
C$^{36}$S, Si$^{34}$S, and Si$^{36}$S, lines and the derived
C$^{34}$S/C$^{36}$S and Si$^{34}$S/Si$^{36}$S intensity ratios are
given in Table\,\ref{table}. We can summarize that we have clearly
detected  C$^{36}$S and Si$^{36}$S in IRC+10216. These are the
first detections of $^{36}$S bearing molecules in a stellar
atmosphere and the first detection of Si$^{36}$S outside the solar
system.
\subsection{The $^{34}$S/$^{36}$S abundance ratio}
The characteristic line profiles of spherically expanding
molecular shells can be explained in terms of line opacity,
expansion velocity and the relative sizes of the telescope beam
and the stellar envelope (e.g. Kuiper et al. \cite{kuiper76};
Olofsson et al. \cite{olofsson82}). The slightly U-shaped profiles
seen in C$^{34}$S and Si$^{34}$S can be explained by optically
thin and spatially resolved emission. Optically thin lines are
expected because the main isotopic lines are only moderately
optically thick and because $^{34}$S is a factor $\sim 20$ less
abundant than the main isotope (Kahane et al. \cite{kahane88};
Cernicharo et al. \cite{cerni2000}). Interferometric observations
(Gu\'elin et al. \cite{guelin93}; Lucas et al. \cite{lucas95}, for
maps see Grewing \cite{grewing94}) confirm a compact condensation
of about 10$''$ diameter for main isotopic CS. SiS is even more
compact (Lucas et al. \cite{lucas95}), which is also apparent by a
comparison of C$^{34}$S and Si$^{34}$S line shapes (Figs.\,
\ref{spectra} and \ref{SiSspectra}). It is safe to assume that
also C$^{36}$S and Si$^{36}$S emission is optically thin and it is
plausible that the distributions of $^{34}$S and $^{36}$S bearing
isotopomers are similar. Chemical fractionation, as discussed in
Chin et al. (\cite{chin96}), should be negligible in the warm
(50\,K, Cernicharo et al. \cite{cerni2000}) environment of the
inner shell of IRC+10216. Since the isotopomers have very similar
transition frequencies we can assume that the intensity ratios of
corresponding transitions reflect the abundance ratios (for a
discussion, see Kahane et al. \cite{kahane88}).

From Table\,\ref{table} it is evident that the relative errors of
the integrated line intensities of the $^{36}$S bearing molecules
are of the order of 20\% or more, while the relative errors for the
$^{34}$S bearing species are much smaller. While it is relatively
straightforward to determine the X$^{36}$S/X$^{34}$S ratios  from
fitted line profiles (as summarized in Table\,\ref{table}), it is
much more difficult to estimate the X$^{34}$S/X$^{36}$S ratios since
the simple rules for error propagation apply only if the relative
errors are small. In addition, if we assume that the error
distribution of the X$^{36}$S/X$^{34}$S ratios is Gaussian, the
distribution of the inverse ratio is not such a simple and symmetric
function, making the determination of the variance difficult (see
Chapter 3.3.3 in Wall \& Jenkins \cite{wall03}).

The errors as quoted have been calculated from the formal errors of
the fits to the line profiles. From these four ratios, and using as
weights the inverse squares of the formal errors of the individual
ratios, we have determined the weighted mean of the
$^{36}$S/$^{34}$S abundance ratio and its error to be 0.0094($\pm
0.0013$) (see e.g. Bevington \& Robinson \cite{bevington92}). From a
graphical representation of the individual values and their mean
(see Fig.\,\ref{abundgraph}) the only transition whose abundance
ratio deviates from the mean value is the SiS $5-4$ line. This might
be an additional hint that for this line there is, in addition to
the statistic uncertainty, also a systematic effect, for example a
blend with an unidentified line (see the discussion above). Since
the relative error of the average $^{36}$S/$^{34}$S abundance ratio
is only 14\%, the expectation value of the $^{34}$S/$^{36}$S ratio
and its error can be easily determined. In the following we will use
a $^{34}$S/$^{36}$S ratio of 107($\pm 15$) for IRC+10216 to ease a
comparison with values published in the literature.

This value of the best fit is very similar to the interstellar
value of 115($\pm17$) (Mauersberger et al. \cite{mauersberger96})
but is smaller than the solar system value of 288 (Ding et al.
\cite{ding2001}). In Table \,\ref{israt}, we give a compilation of
all available estimates for sulfur isotope ratios.
\subsection{The origin of $^{36}$S}

\begin{table*}
\caption[]{\label{israt}Isotopic ratios of sulfur$^{a}$}
\begin{tabular}{c c c c c }
\hline
Ratio&solar & ISM &C.R. &IRC+10216 \\
& system & & & \\
 \hline
$^{32}$S/$^{34}$S&22.64(0.00)&24.4(5)$^b$&16($-$5,+11)&21.8(2.6)\\
$^{32}$S/$^{33}$S&126.95(0.05)&$\sim$150&38($-$18,+500)&121(15)\\
$^{32}$S/$^{36}$S&6519(20)&3280(760)& &2700(600)\\
$^{34}$S/$^{33}$S&5.606(0.002)&6.3(1)&&5.6(.3)\\
$^{34}$S/$^{36}$S&288(1)&115(17)&&107(15)\\
\hline
\end{tabular}
\\ a) References: solar system data: Canyon Diablo troilite (CDT),
Ding et al. \cite{ding2001}; ISM data: Chin et al.
(\cite{chin96}), Mauersberger et al. (\cite{mauersberger96}), most
ISM data are from the inner Galaxy; Cosmic Rays: these are
``source abundances'', Thayer (\cite{thayer97}); IRC+10216: Kahane
et al. (\cite{Kahane00}), this paper. \\b) the average value for
sources at various galactocentric radii; the data by Chin et al.
(\cite{chin96}) indicate a possible galactocentric
$^{32}$S/$^{34}$S abundance gradient, which would correspond to a
value of 32($\pm5$) at the solar circle.
\\
\end{table*}
\subsubsection{Solar abundance and nucleosynthesis of $^{36}$S}
 $^{36}$S  is one of the more enigmatic stable nuclei
in nature. Partly because of its rarity (solar system isotopic
abundance 0.015\% relative to $^{32}$S), and partly because it is
rarely employed in mass spectrometric studies of sulfur isotopes, up
to very recently even its solar system abundance was quite
uncertain. The often used Anders \& Grevesse
(\cite{andersandgrevesse89}) abundance table lists the S isotopes
with a $^{34}$S/$^{36}$S ratio of 211 (no errors given). An updated
value is recommended in the 1998 compilation of isotopic abundances
of the International Union of Pure and Applied Chemistry, (Rosman
{\&} Taylor \cite{rosman1998}), where a ratio $^{34}$S/$^{36}$S =
248$\pm29$ is listed as the ``best measurement from a single
terrestrial source''; this composition has also been used by Lodders
(\cite{lodders03}) for her Solar System abundance table. Sulfur
isotope ratios measured in the laboratory are commonly reported as
permill deviations from the Canyon Diablo Troilite (V-CDT) standard,
which has an S isotopic composition fairly representative for Solar
system materials (IUPAC 2003 compilation; de Laeter et al.,
\cite{delaeter03}). Its composition has recently been precisely
redetermined, with a $^{34}$S/$^{36}$S ratio of 288$\pm 1$ (Ding et
al., \cite{ding2001}). Another modification with respect to Anders
\& Grevesse (\cite{andersandgrevesse89}) comes from a lower sulfur
elemental abundance by a factor 0.86, as first deduced by Palme \&
Beer (\cite{palme93}) and quoted in Lodders (\cite{lodders03}) for
the solar system. All in all, the solar abundance of $^{36}$S is
lower than the commonly used Anders \& Grevesse
(\cite{andersandgrevesse89}) value by a factor 1.6.

\begin{figure}
\includegraphics[width=8.5cm]{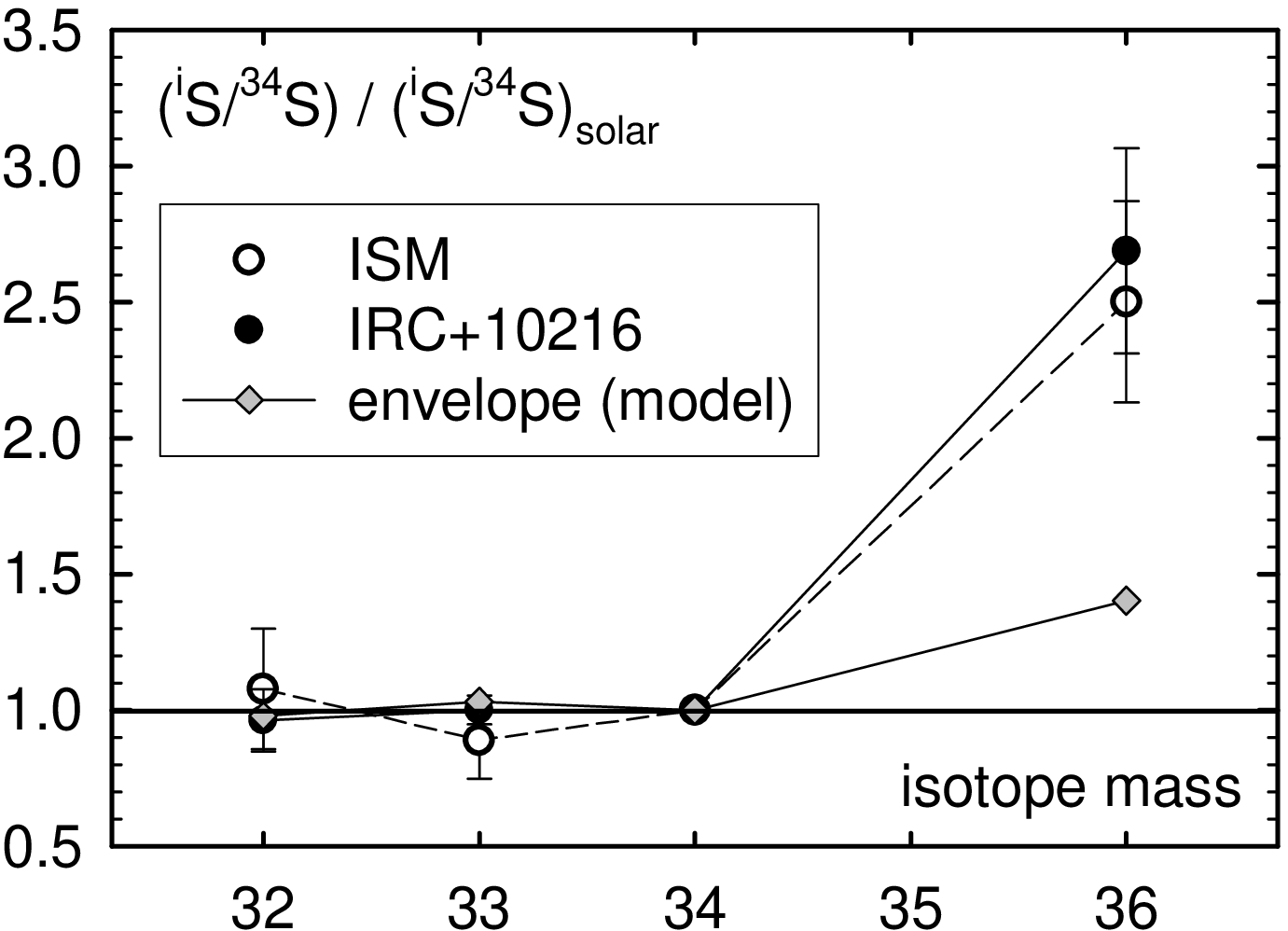}
\caption[]{Sulfur isotopic ratios in IRC+10216, normalized to
$^{34}$S and solar ratios. Data from Kahane et al.
(\cite{Kahane00}) and this work. Also shown are the S isotope
ratios predicted for the envelope of a 1.5\,M$_\odot$
solar-metallicity, TP-AGB star at the time the Cl isotope ratio is
matched (see text; with updated solar abundances), and the ratios
measured for the ISM by Mauersberger et al.
(\cite{mauersberger96}).} \label{abundgraph}
\end{figure}
The former uncertainty on this isotope extends to its
nucleosynthetic origin. The synthesis of the major isotopes of
sulfur, similar as for other elements between Si and Fe, can
mostly be assigned to charged particle reactions during
hydrostatic and explosive burning stages in massive stars:
$^{32}$S and $^{34}$S to hydrostatic and explosive oxygen burning,
$^{33}$S to explosive oxygen and neon burning (Chin et al. 1996;
Woosley et al. 2002). For some exceptionally rare neutron-rich
isotopes in this mass range including $^{36}$S, however, the
situation may be different. $^{36}$S has a closed neutron shell,
hence a small neutron capture cross section, and in an s-process
it could, in principle, act as a bottleneck building up a
significant abundance relative to neighboring nuclides. For this
reason, the {\em weak s-process} taking place during core He
burning and convective shell C (Ne) burning in massive stars has
been suggested as a major source of the $^{36}$S abundance in the
Solar System (Schatz et al. \cite{schatz95}; Woosley et al.
\cite{woosley02}), and according to first network calculations by
Schatz et al. (\cite{schatz95}) it seemed to be able to fulfill
that role. However, re-assessment by Reifarth et al.
(\cite{reifarth2000}) with updated nuclear input data indicates a
more complex situation. The unexpectedly low neutron capture cross
section of $^{34}$S causes this nuclide to restrict the flow
towards $^{36}$S. With rare $^{35}$Cl being the main seed nucleus
rather than abundant $^{28}$Si and $^{32}$S, the weak s-process,
while actually making $^{36}$S, falls short by more than an order
of magnitude in quantity in their assessment. Here, a further
complication arises because in all recent calculations the
abundance of the major seed, $^{35}$Cl, which in the s-process
zone is given by its original abundance, has been taken from
Table\,3 of Anders and Grevesse (\cite{andersandgrevesse89}).
Actually, there is disagreement for Cl between the nuclide
abundances in their Table 3 and the 39{\%} higher elemental
abundance given in their Table\,1; the latter agrees with other
previous and more recent compilations. The predicted
$^{35}$Cl/$^{37}$Cl ratio is not affected by these changes, but
there is a significant effect on the $^{36}$S overproduction
factor, which is increased by a factor of 1.4.

According to current models, in massive stars most $^{36}$S is
produced in a large mass zone where convective shell C-burning
operates in hydrostatic conditions before the SN\,II explosion
(Woosley \& Weaver \cite{woosley1995}). Owing to the high neutron
density in the first phase of carbon consumption (Raiteri et al.
\cite{raiteri92}), more $^{36}$S is fed through the neutron channel
of the unstable $^{35}$S than would be inferred by the occurence of
the classical weak s-process. The nucleosynthesis yields of the most
recent massive star calculations, made with solar initial
composition and with a full network included (Woosley et al.
\cite{woosley02}; Rauscher et al. \cite{rauscher02}; Limongi \&
Chieffi \cite{Limongi03}), are all based on the Bao et al.
(\cite{bao2000}) compilation of neutron capture cross sections
(where the new and much lower cross section for $^{34}$S is
included). These results show a production factor of $^{36}$S that
is typically low by a factor of 2 with respect to the mean of the
most abundant nuclides. Taken at face value, this factor of 2 may be
reconciled using the new solar $^{36}$S abundance as well as the
correct $^{35}$Cl abundance. However, this does not completely solve
the problem of the reproduction of solar $^{36}$S, since one would
expect the weak s-process products to decline with decreasing
metallicity, i.e. earlier generations of massive stars to have been
less effective producers of $^{36}$S. In addition, there are
uncertainties related to the nuclear input data. In particular, the
rate of the major channel $^{36}$Cl(n,p)$^{36}$S may be subject to
further refinements in\ the future owing to the much better energy
resolution obtained in the new measurements of Wagemans et al.
(\cite{wagemans}). On the other hand, recent investigations of
n-induced reactions on $^{37}$Ar and $^{39}$Ar (Goeminne et al.
\cite{Goeminne2000}, \cite{goeminne2002}) seem not to affect the
$^{36}$S production in a significant manner.

A complementary, though probably less important, source to the
weak s-process is the {\em main s-process} taking place in the He
burning shell of low mass AGB stars. According to calculations
partially published in Travaglio et al. (\cite{travaglio04}), the
Galactic contribution to $^{36}$S at the epoch of Solar System
formation by all previous generations of AGB stars is about 4{\%};
this may reach $\sim $10{\%} once the revised lower $^{36}$S and
higher $^{35}$Cl solar abundances are taken into account.
\subsubsection{Stellar mass of IRC+10216}
In Fig.\,\ref{abundgraph} we show the S isotopic data for
IRC+10216 (from this work and Kahane et al. \cite{Kahane00})
normalized to $^{34}$S and solar system ratios. The data clearly
show a more than a factor of 2 enhancement of $^{36}$S/$^{34}$S in
IRC+10216 while the other ratios are solar within error. An
s-process enhancement of $^{37}$Cl in IRC+10216 was found by
Kahane et al. (\cite{Kahane00}). These authors were able to match
the observation with the predictions of a low mass
(1.5\,M$_{\odot})$ AGB star model of solar metallicity. The
enhancement predicted for the envelope after the 15th thermal
pulse just corresponded to the observed Cl isotope ratio.
Re-analysis, using the updated solar abundances, confirms the
agreement, which persists for the last three thermal pulses with
dredge-up ({\#}15 to {\#}17). Included in Fig.\,\ref{abundgraph}
are the corresponding predictions for the sulfur isotopes
(assuming, as in the case of Cl, the updated solar starting
composition). It is obvious in Fig.\,\ref{abundgraph} that
qualitatively there is agreement between model and observations in
that in IRC+10216 an enhancement of only $^{36}$S/$^{34}$S is
seen. However, our observed enhancement of $\sim $170{\%} in this
ratio is more than a factor of 4 higher than even the new
prediction, which already is about twice as high ($\sim $40 {\%}
instead of $\sim $20 {\%}) than that in Kahane et al.
(\cite{Kahane00}).

It has been argued in the past that the central star of IRC+10216
might correspond to an AGB star of intermediate mass, of around
5\,M$_\odot$, instead of the low mass model discussed here. This
possibility was at the upper limit of the range of uncertainty of
the available distance estimates of 110 --- 170 pc (Winters et al.
\cite{winters94}; Crosas \& Menten \cite{Crosas97}; Le Bertre
\cite{lebertre97}; Groenewegen et al. \cite{groenewegen98}; Weigelt
et al. \cite{weigelt98}). An intermediate mass was inferred by
Gu\'elin et al. (\cite{guelin95}) on the basis of older estimates of
some isotope ratios (see the discussion in Kahane et al.
\cite{Kahane00}). We have computed a series of AGB models of
different initial mass using updated network of neutron capture
cross sections and solar abundances. These confirm the conclusions
already reached by Kahane et al. (\cite{Kahane00}) that an
intermediate mass of 5\,M$_\odot$ is excluded by the almost solar Mg
isotope ratios observed. Indeed, in intermediate mass stars the
$^{22}$Ne($\alpha$,n)$^{25}$Mg and
$^{22}$Ne($\alpha$,$\gamma$)$^{26}$Mg reactions would be more
efficiently activated by the higher temperature in the convective He
flashes, providing a clear excess of both neutron-rich Mg isotopes.
Supporting evidence derives from the observed $^{37}$Cl/$^{35}$Cl
ratio, which would be much higher in an intermediate mass star. The
resulting $^{34}$S/$^{36}$S ratio, in any case, does not depend much
on the initial mass.

While the enhancement of the observed $^{36}$S/$^{34}$S ratio with
respect to the 1.5\,M$_\odot$ models could, in principle, be due
to some unrecognized problems with the AGB models or assumptions
on nuclear parameters involved with it, we believe this to be
highly unlikely and the main s-process in IRC+10216 to be not the
main source of its (relative to solar) enhanced $^{36}$S/$^{34}$S
ratio. Alternative explanations are briefly discussed below.
\subsubsection{Chemical evolution}
One clue may be the close agreement between the value for IRC+10216
and the one reported for the ISM (also shown in
Fig.\,\ref{abundgraph}) by Mauersberger et al.
(\cite{mauersberger96}), which at first glance is puzzling. A
possible solution may be afforded by Galactic chemical evolution
(GCE) and the different times when IRC+10216 and the Solar System
formed, i.e. assuming a starting composition different from solar.
Based mostly on the $^{12}$C/$^{13}$C ratio and limits to C/O,
Kahane et al. (\cite{Kahane00}) derived an upper limit of 2
M$_{\odot}$ for IRC+10216, which nevertheless still allows the
object to be as young as $\sim $1.5 Ga. Assuming a correspondingly
evolved starting composition, observations might then be compatible
(within errors) with model predictions. Indeed, first, there is no
need for a large intrinsic contribution to $^{36}$S from IRC+10216
(in agreement with the relatively small effect predicted by the AGB
star model; Fig.\,\ref{abundgraph}). And second, the large
difference between the current ISM (and IRC+10216) and the Solar
System, is compatible with massive star model predictions, where the
$^{36}$S Galactic contribution increases with metallicity (as
already pointed out by Mauersberger et al., \cite{mauersberger96}).
This would require a rather different evolution of the Cl and S
isotopic ratios during the last few Ga, since the observed
$^{35}$Cl/$^{37}$Cl ratio in IRC+10216 is matched by AGB model
predictions using solar as the starting composition. Indeed, both Cl
isotopes are thought to be synthesized in massive stars in a primary
way during explosive Ne and O burning (Weaver \& Woosley 1995). Only
4\% of the solar abundance of $^{37}$Cl is contributed by the main
s-component in low mass AGB stars, while $^{35}$Cl gets actually
depleted by neutron captures, according to the GCE calculations by
Travaglio et al. (\cite{travaglio04}). On the observational side,
$^{35}$Cl/$^{37}$Cl in Orion has been found to be enhanced relative
to solar (Salez et al., \cite{salez96}), but the errors are too
large to reach a firm conclusion.

A further possibility is that IRC+10216 simply started with a
$^{36}$S abundance that was unusually high for its time of
formation and location, i.e. invoke inhomogeneity in the ISM. Of
course, it could as well be the Solar System rather than IRC+10216
that started out with an unusual composition. Both suggestions are
not purely ad hoc. Already Lugaro et al. (\cite{Lugaro99}), in
order to explain isotopic variations in the Si isotopes of single
presolar silicon carbide grains from primitive meteorites, called
for small local inhomogeneities in the ISM at the time of parent
star formation of the grains as an alternative to Galactic
chemical evolution. And, as has been known for quite some time
already, oxygen isotopic ratios in the Solar System seem to be
quite peculiar (e.g., Wilson \& Rood, \cite{wilson94}; Prantzos et
al., \cite{prantzos96}). In addition, the interstellar Rb isotope
ratio $^{85}$Rb/$^{87}$Rb determined toward $\rho$ Ophiuchi A
(Federman et al. \cite{federman04}) indicates a higher ratio of r-
to s-process nuclides in the Solar System as compared to the local
ISM.

Reifarth et al. (\cite{reifarth2000}) discussed various other
possible nucleosynthesis scenarios besides the s-process.
Accepting Galactic chemical evolution as the reason for the
difference between the Solar System and the ISM $^{34}$S/$^{36}$S
ratios, those that involve core collapse supernovae (SN) (other
than the s-process) face the problem of how to account for the
large increase during the last 4.6\,Ga. This probably includes the
r-process. To some extent this also includes the ``neutron
burst'', intermediate between s- and r-process and thought to
occur during passage of a shock wave through the He shell of type
II SN, which Meyer et al. (\cite{meyer2000}) modelled to explain
specific types of isotope abundance anomalies found in supernova
grains preserved in meteorites. Note, however, that the efficiency
of the process depends on previous s-process seeds and that its
effects are included in the SN\,II yields calculated by the
authors cited above. Low-entropy scenarios associated with
supernovae type Ia that can account for production of neutron-rich
isotopes such as $^{48}$Ca and $^{50}$Ti (Meyer et al.
\cite{meyer96}; Kratz et al. \cite{Kratz2001}) would be
qualitatively compatible with the trend in GCE, but again existing
information on production yields for $^{36}$S is not encouraging
(Reifarth et al. \cite{reifarth2000}; Kratz et al.
\cite{Kratz2001}).

In summary, while it appears we have observationally identified a
stellar source with enhanced $^{36}$S and there are some
reasonable ideas about how Galactic $^{36}$S is produced, we are
left with a number of open questions. Further studies of other
objects and/or regions in the ISM, especially when correlated with
other isotopic systems such as the carbon, oxygen, and chlorine
isotopes, may turn out to be helpful. The same holds true for
continued study of the relevant stellar physics and
nucleosynthesis processes.

\section{Conclusions}
The main conclusions of this paper are:
   \begin{enumerate}
      \item Measuring the rare isotopic species of CS and SiS in IRC+10216, the abundance
   ratio of $^{34}$S/$^{36}$S in IRC+10216 is 107($\pm 15$), comparable to values in the Galactic
   interstellar medium (115), but smaller than the solar value
   (288).
      \item
      The increase of the
   $^{36}$S abundance relative to $^{34}$S only qualitatively
   follows the prediction of a low mass AGB star model. Quantitative agreement of the observed $^{34}$S/$^{36}$S ratio
   with the stellar models can be reached if the age of IRC+10216 and Galactic Chemical evolution are taken into account
   or if there are considerable inhomogeneities in the interstellar medium, and either
   IRC+10216 or the Sun started with an unusual $^{36}$S abundance.
   \item
    From the observed line density toward IRC+10216 and toward Galactic star forming regions,
   confusion limits toward these sources are estimated.

 \end{enumerate}

\appendix
\section{Line identifications}
\label{appa}
\begin{table}
\caption[]{\label{lineidentification} Line identifications}
\begin{tabular}{r r r r r }
\hline Nr.& $T_{\rm A}^*$ & Freq.$^a$ & Im.-Freq$^a$ & Identification\\
& mK & MHz & MHz\\ \hline 1 & 4 &{\bf94903} & 98126 &HCCC$^{13}$CN $36-35$  \\
2 & 3.5 &94913& {\bf98155}&SiC$_4$ $32-31$ (im.)\\
3 & 5.5 &95018&{\bf98011} &l-C$_3$H$^b$ (im.)\\
4 & 6 & 95035 & {\bf 97095} &C$^{36}$S; l-C$_3$H$^b$ (im.)\\
5 & 21 & 95049&{\bf 97980} & CS(2-1) (im.) \\
6 & 20 & {\bf 95087}&97942 & SiC4 $31-30$\\
7 & 1.5 & {\bf U95115}& 97914 & \\
8 & 1.5 & {\bf 95126}& 97903 & SiCCC$^{13}$C $32-31$\\
\\
9 & 20 & {\bf 142297} & 151263 &H$_2$C$_4$ $16_{1,16}-15_{2,15}$\\
10& 23&{\bf 142321} &151235 & Al$^{37}$Cl $10-9^c$\\
11& 3.5& {\bf 142345}& 151215& HCCC$^{13}$CN $54-53$\\
12& 5& {\bf 142402}& 151158& SiS $J=8-7, v=4^d$?\\
13&29 & {\bf 142410}& 151150& NaCN $9_{27}-8_{26}^c$\\
14& 5& {\bf U142489}  &151071 & \\
15& 40& {\bf 142502}& 151058& CCS $11_{11}-10_{10}^c$\\
16& 7&{\bf 142525} &151035 & C$^{36}$S $3-2^c$\\
17& 460& {\bf 142560}& 151000& $^{29}$SiS $8-7^c$\\
18& 4&{\bf U142594} &150966 & \\
19& 12&{\bf U142622} &150940 & \\
20& 7&{\bf 142675} &150885 & SiCCC$^{13}$C $48-47$\\
21& 9&142695&150865  & artefact\\
22& 850&{\bf 142730} &150831 &C$_4$H $N=15-14^c$ \\
23& 9& {\bf 142755}&150805 &C$_4$H $N=15-14$? \\
24& 850& {\bf 142768}&150793 &C$_4$H $N=15-14^c$\\
 \hline
\end{tabular}\\
\begin{footnotesize}
a) The band from which the line probably originates is given in
bold face; for information, the frequency of the corresponding
image band is also given.  b) Thaddeus et al. (\cite{thaddeus85}).
c) Cernicharo et al. (\cite{cerni2000}) and references therein. d)
The SiS $v=4$ $J=8-7$ line is at 142399.8 MHz. A shift of $1-2$
MHz could be possible if the
line is produced in the acceleration region (blue).\\
\end{footnotesize}
\end{table} In
Table\,\ref{lineidentification} we list the observed spectral line
features seen in Fig.\,\ref{spectraall}, and the approximate
values for $T_{\rm A}^*$, the rest frequency and the rest
frequency in the image band (assuming an LSR velocity of $-26\,
\rm km\,s^{-1}$). The final column contains some identifications
from the literature. In the 256 MHz wide spectrum of the C$^{36}$S
$2-1$ line we have identified 8 spectral line features. Of those,
three can be identified with transitions from the image sideband.
In the 256 MHz around the frequency of the C$^{36}$S $3-2$ line we
have detected 16 spectral line features, none of which can be
assigned to a line from the image band. Six lines can be
identified with known transitions from the Cernicharo et al.
(\cite{cerni2000}) line survey, and the other identifications were
made with the catalog of the Instituto de la Estructura de la
Materia (Madrid) containing $\sim 1000$ species described by the
authors. There is a clearly detected feature at the frequency of
the C$^{36}$S line; another feature at 142.755 GHz is tentatively
assigned to emission from C$_4$S. Eight  further lines have not
been identified. The emission at 142.695\,GHz is an artefact due
to some bad spectrometer channels.

\section{The confusion limit toward astronomical molecular line sources}
\label{appb}
\begin{table}
\caption[]{The confusion limit reached by the IRAM 30-m telescope
toward some molecular line sources}
\begin{tabular}{r r r r  r}
\hline Source &$\lambda$& weakest & $t_{\rm int}^{a)}$ & Ref.$^b$\\
$\Delta v$&&detect. Lines& & \\
&mm&mK&  h & \\
\hline IRC+10216 & 3 & 1.5 &  28&1)\\
28\,km\,s$^{-1}$&2&3.5&10 &1, 2) \\
& 1.3 & $\la 2$& 80&2)\\
\hline\\
 Orion-KL& 3 & 20 & 0.8&3)\\
5\,km\,s$^{-1}$& 2 & 50 &0.3 &4) \\
&1.3&70 &0.4&3)\\
\hline SgrB2(N)& 3 & $\sim 20$ & 0.4& 3)\\
$\sim$ 17\,km\,s$^{-1}$& 2 & $\sim 30$ & 0.4 &3) \\
& 1.3& $\sim 40$  & 0.5&3) \\
\end{tabular}
\begin{footnotesize}\\
a) integration time (on+off) with present receivers at the IRAM
30-m telescope under normal winter conditions (good summer
conditions), one polarization to obtain an rms of 1/3 $T_{\rm mb}$
with a velocity resolution of 1/5 the line width (one polarization
receiver only);  b)References: 1) this paper 2) Ziurys et al.
(\cite{ziurys2002}); 3) unpublished data; 4) Mauersberger et al.
(\cite{mau88});  c) weaker lines can be identified if one makes
use of the unique line shapes of the spectra of this source
(Cernicharo et al. \cite{cerni2004}).
\end{footnotesize}
\end{table}
It is evident from Sect.\,\ref{confusionlim} that in our 2 and
3\,mm spectra of IRC+10216 we are close to the confusion limit,
where an increase of integration time does not yield much further
information. It is interesting to investigate where line confusion
begins to play a role for other favorites of molecule hunters such
as the Orion Hot Core, Sgr\,B2 or the starburst galaxy NGC\,253.
We limit our discussion to the IRAM-30m telescope. If one wants to
extrapolate our results to other telescopes with a higher or lower
resolution one has to take into account the detailed source
structure chemical and physical conditions within the regions
observed (Comito \& Schilke \cite{comito02}).

The definition of the useful observing time or rms to be reached
is by no means unique and depends on whether one is interested in
a mere detection of a line or whether one also wants to obtain
some detailed information on the line shape. Here we try to be
pragmatic: we have investigated  spectra of several molecular line
sources made with the IRAM 30-m telescope at 3\,mm, 2\,mm and
1.3\,mm wavelength made with such a long integration that $\ga
50\%$ of the spectral range observed is covered with features. We
have determined the antenna temperature of the weakest unequivocal
line features (knowing the typical line width and shape in the
sources studied). The results are given in
Table\,\ref{confusionlim}. We also give an indication of the
necessary rms for a 3\,$\sigma$ detection with a velocity
resolution 1/5 the full line widths typical for these sources and
the corresponding necessary integration time with the 30-m
telescope and its present receivers (including all observing
overheads, using one polarization only). In Orion, part of the
confusion arises because of the large line widths observed in the
outflow source. This can e.g. be prevented by observing at a
carefully selected position offset from the outflow source (Combes
et al. \cite{combes96}).
\begin{acknowledgements}

      V. Ahrens from the I. Physikalisches Institut
      der Universit\"at K\"oln gave us fits to the line
      frequencies of C$^{36}$S prior to publication. R.G.
      acknowledges support by the Italian MIUR-Cofin2002 Project
      ``Nucleosynthesis in the Early Phases of the Universe''.

\end{acknowledgements}

\end{document}